\documentclass[aps,showpacs,preprint,superscriptaddress]{revtex4}

\usepackage{amsmath,amssymb,amsfonts}
\usepackage[english]{babel}
\usepackage{graphicx}

\begin{document}

\title{Operator methods applied to special functions}

\author{H. Moya-Cessa and F. Soto-Eguibar}
\affiliation{Instituto Nacional de Astrof\'isica \'Optica y Electr\'onica\\
Calle Luis Enrique Erro No. 1, Sta. Ma. Tonantzintla, Pue. CP 72840, Mexico}
\begin{abstract}
Based on operator algebras commonly used in quantum
mechanics some properties of special functions such as Hermite  and Laguerre polynomials and Bessel functions are derived.
\end{abstract}
\maketitle
\section{Introduction}
Based on some techniques that are common to quantum mechanics, we give some examples on how some series of special functions may be added.
For this, we use some theorems and lemmas that are not usually known when one studies special functions, such as the Baker-Hausdorff formula and the Hadamard lemma.
We developed further preliminary results presented in  \cite{arXive}.

\section{Hermite polynomials}
The generating function for the Hermite polynomials is
\begin{equation}
e^{-\alpha ^{2}+2\alpha x}=\sum\limits_{n=0}^{\infty}H_{n}(x)\frac{\alpha^{n}}{n!}.
\end{equation}

The Hermite polynomials may be obtained from Rodrigues' formula \cite{A}
\begin{equation}\label{1q}
H_{n}(x)=(-1)^{n}e^{x^{2}}\frac{d^{n}}{dx^{n}}e^{-x^{2}}.
\end{equation}
From the recurrence relations
\begin{equation}
H_{n+1}(x)=2xH_n(x)-2nH_{n-1}(x),
\end{equation}
and
\begin{equation}
\frac{dH_n(x)}{dx}=2nH_{n-1}(x),
\end{equation}
we can generate all the Hermite polynomials. \\

 From the above recurrence relations, we can also prove that, if we define the functions
\begin{equation}\label{fooa}
\psi_n(x)=\frac{\pi^{-1/4}}{\sqrt{2^{n}n!}}\mathrm{e}^{-x^2/2}H_n(x),
\end{equation}
then
\begin{equation}
A^{\dagger}\psi_n(x)\equiv\frac{1}{\sqrt{2}}\left(x-\frac{d}{dx}\right)\psi_n(x)=\sqrt{n+1}\psi_{n+1}(x)
\end{equation}
and
\begin{equation}
A\psi_n(x)\equiv\frac{1}{\sqrt{2}}\left(x+\frac{d}{dx}\right)\psi_n(x)=\sqrt{n}\psi_{n-1}(x).
\end{equation}
The functions (\ref{fooa}) constitute a complete orthonormal set for the space of square integrable functions, then we can expand any function in that space as
\begin{equation}\label{2.9}
f(x)=\sum_{n=0}^{\infty}c_n \psi_n(x),
\end{equation}
where
\begin{equation}
c_n =\int_{-\infty}^{\infty}dxf(x)\psi_n(x).
\end{equation}

Hermite polynomials are also solutions of the second order ordinary differential equation
\begin{equation}
y''-2xy'+2ny=0.
\end{equation}

Let us define the differential operator $p$ as
\begin{equation}\label{p}
p=-i\frac{d}{dx},
\end{equation}
then we have that
\begin{equation}
\frac{d^{n}}{dx^{n}}=\left( -\frac{p}{i}\right) ^{n},
\end{equation}
and we can rewrite (\ref{1q}) in the form
\begin{equation}\label{herop}
H_{n}(x)=(-i)^{n}e^{x^{2}}p^{n}e^{-x^{2}}=(-i)^{n}\left(e^{x^{2}}pe^{-x^{2}}\right) ^{n}.
\end{equation}
The operator inside the parenthesis above has the form
\begin{equation}
e^{\xi A}Be^{-\xi A},
\end{equation}
for which we will use the following lemma \cite{L,Soto}.
\begin{quote}
\textbf{Hadamard lemma:} Given two linear operators $A$ and $B$ then
\begin{eqnarray}   \label{bch0}
e^{\xi A}Be^{-\xi A}=B+\xi\left[ A,B\right]+\frac{\xi^{2}}{2!}\left[ A,\left[A,B\right]\right]+\frac{\xi^{3}}{3!}\left[A,\left[A,\left[A,B\right]\right]\right]+\cdots,
\end{eqnarray}
where $[A,B]\equiv AB-BA$ is the commutator of operators $A$ and $B$. \\
\end{quote}

This allows us  to obtain an expression for
\begin{equation}
H_{n}(x)=(-i)^{n}\left(e^{x^{2}}pe^{-x^{2}}\right) ^{n},
\end{equation}
formula (\ref{herop}) developed above.\\
We identify
\begin{equation}
\xi =1,\qquad A=x^{2},\qquad B=p,\nonumber\\
\end{equation}
in equation  (\ref{bch0}), so that
\begin{equation}\label{formop}
e^{x^{2}}pe^{-x^{2}}
=p+1\left[ x^{2},p\right]+\frac{1}{2!}\left[ x^{2},\left[x^{2},p\right]\right]+\frac{1}{3!}\left[x^{2},\left[x^{2},\left[x^{2},p\right]\right]\right]+\cdots
\end{equation}
To calculate the first commutator $\left[ x^{2},p\right]$, we use the general property
\begin{equation}\label{p1c}
[AB,C]=A[B,C]+[A,C]B
\end{equation}
of commutators, and that
\begin{equation}
[x,p]f(x)\equiv
-i(x\frac{d}{dx}-\frac{d}{dx}x)f(x)=-xf'(x)+xf'(x)+if(x)=if(x);
\end{equation}
i.e. $[x,p]=i$, to get the commutation relation
\begin{equation}
\left[ x^{2},p\right] =2ix.
\end{equation}
It is obvious that all the other commutators in (\ref{formop}) are zero, and we finally get
\begin{equation}\label{Hermite}
H_{n}(x)=(-i)^{n}\left( p+2ix\right)^{n}1 .
\end{equation}
This last expression can be used to obtain the generating function. We have,
\begin{equation}\label{2q}
\sum_{n=0}^{\infty }H_{n}(x)\frac{\alpha ^{n}}{n!}
=\sum_{n=0}^{\infty }\frac{\alpha ^{n}}{n!}(-i)^{n}\left(
p+2ix\right) ^{n}1=e^{-i\alpha \left( p+2ix\right) } 1.
\end{equation}
We obtained the exponential of the sum of two quantities that do not commute.
The above exponential can be factorized in the product of
exponentials via the Baker-Hausdorff formula
\subsection{ Baker-Hausdorff formula}
\begin{quote}
\textbf{Baker-Hausdorff formula:} \textit{Given two operators} $A$
\textit{and} $B$ \textit{that obey}
\begin{equation}
\left[ \left[ A,B\right] ,A\right] =\left[ \left[ A,B\right],B\right] =0,
\end{equation}
\textit{then}
\begin{equation}\label{bhformula}
e^{A+B}=e^{-\frac{1}{2}\left[ A,B\right]}e^{A}e^{B}.
\end{equation}
\end{quote}

Then we  identify
\begin{equation}
A=2\alpha x,
\end{equation}
\begin{equation}
B=-i\alpha p,
\end{equation}
and then
\begin{equation}
\left[ A,B\right] =-2i\alpha ^{2}\left[ x,p\right] =2\alpha ^{2},
\end{equation}
such that
\begin{equation}
\sum_{n=0}^{\infty }H_{n}(x)\frac{\alpha ^{n}}{n!}=e^{-\alpha^{2}}e^{2\alpha x}e^{-i\alpha p}1.
\end{equation}
Using now the obvious fact that $e^{-i\alpha p}1=1$, we finally obtain
\begin{equation}
\sum_{n=0}^{\infty }H_{n}(x)\frac{\alpha ^{n}}{n!}=e^{-\alpha ^{2}+2\alpha x},
\end{equation}
that is the generating function for Hermite polynomials. \\

\subsection{Series of even Hermite polynomials}
In order to show the power of the operator methods, we calculate now the value of the following even Hermite polynomials series,
\begin{equation}
F(t)=\sum\limits_{n=0}^{\infty }\frac{t^{n}}{n!}H_{2n}\left( x\right).
\end{equation}
 From (\ref{Hermite}), we get
\begin{equation}
H_{2n}\left( x\right) =\left( -1\right) ^{n}\left( p+2ix\right)^{2n}1.
\end{equation}
Therefore,
\begin{eqnarray}    \nonumber
F(t)&=&\sum\limits_{n=0}^{\infty }\frac{t^{n}}{n!}H_{2n}\left( x\right)
=\sum\limits_{n=0}^{\infty }\frac{t^{n}}{n!}\left( -1\right)
^{n}\left( p+2ix\right) ^{2n}1 \\
&=& \sum\limits_{n=0}^{\infty
}\frac{t^{n}}{n!}\left( -1\right) ^{n}\left[ \left( p+2ix\right)
^{2}\right] ^{n}1=\exp \left[ -t\left( p+2ix\right) ^{2}\right] 1.
 \end{eqnarray}
Developing the power in the exponential above we get
\begin{equation}\label{211000}
F\left(t\right)=\sum\limits_{n=0}^{\infty}\frac{t^{n}}{n!}H_{2n}\left(x\right)=\exp\left\{-t\left[p^{2}-4x^{2}+2i\left(xp+px\right)\right]\right\}1.
\end{equation}
The operators in the exponential in this last expression do not satisfy the conditions of the Baker-Hausdorff formula, so we need another method to understand the action of the full operator that appears in the right side of expression (\ref{211000}). What we do is to propose the ansatz,
\begin{equation}\label{ansatz}
{F\left(t\right)=\exp\left[f\left( t\right)x^{2}\right]\exp\left[g\left(t\right)\left(xp+px\right)\right]\exp\left[h\left(t\right)p^{2}\right]1},
\end{equation}
where $f(t), g(t)$ and $h(t)$ are functions to be  determined.
Deriving this expression with respect to $t$, and dropping the explicit dependence of $f(t), g(t)$ and $h(t)$ on $t$,
\begin{eqnarray}\label{44}\nonumber
&&\frac{dF\left(t\right)}{dt}= \\
&=&\frac{df}{dt}x^{2}F\left(t\right)+\frac{dg}{dt}\exp\left(fx^{2}\right)\left( xp+px\right)\exp\left[g\left( xp+px\right) \right] \exp\left(hp^{2}\right)1\nonumber\\
&+&\frac{dh}{dt}\exp\left(fx^{2}\right)\exp\left[g\left(xp+px\right)\right]p^{2}\exp\left(hp^{2}\right)1.
\end{eqnarray}
Introducing an "smart" 1 in the second and third term, we get
\begin{eqnarray}  \label{44}\nonumber
\frac{dF\left(t\right)}{dt}&=&\frac{df}{dt}x^{2}F\left(t\right)+\frac{dg}{dt}e^{fx^{2}}\left( xp+px\right)e^{-fx^{2}}F(t)  \\
&+&\frac{dh}{dt}e^{fx^{2}} \exp \left[ g\left( xp+px\right) \right] p^{2}\exp\left[ -g\left( xp+px\right) \right] e^{-fx^{2}}F(t).
\end{eqnarray}
We work then with the operator in the second term; we have to use the very useful expression
\begin{eqnarray}\nonumber
e^{\xi A}Be^{-\xi A}=B+\xi \left[ A,B\right] +\frac{\xi ^{2}}{2!}\left[ A,%
\left[ A,B\right] \right] +\frac{\xi ^{3}}{3!}\left[ A,\left[ A,\left[ A,B\right] \right] \right] +\cdots,
\end{eqnarray}
and obtain
\begin{eqnarray} \label{48}
&&\exp \left( fx^{2}\right) \left( xp+px\right) \exp \left(-fx^{2}\right)=xp+px+f\left[ x^{2},xp+px\right] \\
&+&\frac{f^{2}}{2!}\left[ x^{2},\left[ x^{2},xp+px\right] \right]+\frac{f^{3}}{3!}\left[ x^{2},\left[ x^{2},\left[x^{2},xp+px\right] \right] \right]+\cdots.
\end{eqnarray}
The first commutator that appears in the above expression is easily calculated,
\begin{equation}
[x^{2},xp+px] =4ix^{2},
\end{equation}
and so all the others commutators are zero. Substituting back in (\ref{48}), we get
\begin{equation}
\exp \left( fx^{2}\right) \left( xp+px\right) \exp \left(-fx^{2}\right) =xp+px+4ifx^{2}.
\end{equation}
We analyze now the third operator in expression (\ref{44}), i.e.
\begin{equation}
\exp(-fx^{2})\exp \left[ g\left( xp+px\right) \right] p^{2}\exp\left[ -g\left( xp+px\right) \right]\exp(-fx^{2}).
\end{equation}
We study first only a part of it, $\exp \left[ g\left( xp+px\right) \right] p^{2}\exp \left[-g\left(xp+px\right) \right]$. Using again (\ref{bch0}),
\begin{eqnarray}\nonumber
&&\exp \left[ g\left( xp+px\right) \right] p^{2}\exp \left[-g\left(xp+px\right) \right] = \\
&&p^{2}+g\left[ xp+px,p^{2}\right] +\frac{g^{2}}{2!}\left[ xp+px,\left[ xp+px,p^{2}\right] \right] \\
&+&\frac{g^{3}}{3!}\left[ xp+px,\left[ xp+px,\left[xp+px,p^{2}\right] \right] \right] +\cdots.
\end{eqnarray}
Calculating the first commutators,
\begin{equation}
\left[ xp+px,p^{2}\right] =4ip^{2},
\end{equation}
\begin{equation}
\left[ xp+px,\left[ xp+px,p^{2}\right]\right] =-16p^{2},
\end{equation}
\begin{equation}
\left[ xp+px,\left[xp+px,\left[ xp+px,p^{2}\right] \right] \right] =-64ip^{2},
\end{equation}
and so on. It is clear that
\begin{equation}
\exp \left[ g\left( xp+px\right) \right] p^{2}\exp \left[-g\left(
xp+px\right) \right] =p^{2}\sum\limits_{j=0}^{\infty }\left( 4i\right) ^{j}
\frac{g^{j}}{j!}=p^{2}\exp \left( 4ig\right).
\end{equation}
We proceed now to complete the study of the third operator in expression (\ref{44}). Until now we have
\begin{eqnarray}\nonumber
&&\exp(-fx^{2})\exp \left[ g\left( xp+px\right) \right] p^{2}\exp\left[ -g\left( xp+px\right) \right]\exp(-fx^{2})= \\
&&\exp(-fx^{2}) p^{2}\exp \left( 4ig\right) \exp(-fx^{2}).
 \end{eqnarray}
We use once more formula (\ref{bch0}), to write
\begin{eqnarray}\nonumber
&&\exp \left( fx^{2}\right) p^{2}\exp \left( -fx^{2}\right)= \\
&&p^{2}+f\left[x^{2},p^{2}\right] +\frac{f^{2}}{2!}\left[ x^{2},\left[ x^{2},p^{2}\right]
\right] +\frac{f^{3}}{3!}\left[ x^{2},\left[ x^{2},\left[x^{2},p^{2}\right] \right] \right] + \cdots
 \end{eqnarray}
The first commutator gives
\begin{equation}
\left[ x^{2},p^{2}\right] =-2+4ipx,
\end{equation}
the second one gives
\begin{equation}
\left[x^{2},\left[ x^{2},p^{2}\right] \right] =-8x^{2},
\end{equation}
and the third one
\begin{equation}
\left[ x^{2},\left[ x^{2},\left[ x^{2},p^{2}\right] \right]\right] =0;
\end{equation}
such that all the other commutators are zero, and
\begin{eqnarray}\nonumber
&&\exp \left( fx^{2}\right) p^{2}\exp \left( -fx^{2}\right)= \\
&&p^{2}+f\left( -2+4ipx\right) +\frac{f^{2}}{2!}\left(-8x^{2}\right)
=p^{2}+2if\left( xp+px\right) -4f^{2}x^{2}.
  \end{eqnarray}
Finally, we can write a reduced expression for the derivative of the original series $F(t)$,
\begin{eqnarray}
&&\frac{dF\left( t\right) }{dt}=
\allowbreak \Big\{ \frac{df}{dt}x^{2}+\frac{dg}{dt}\left( xp+px+4ifx^{2}\right) + \\
&&\exp \left( 4ig\right)\frac{dh}{dt}\left[ p^{2}+2if\left( xp+px\right)-4f^{2}x^{2}\right] \Big\} F(t)
   \end{eqnarray}
and rearranging terms
\begin{eqnarray}\nonumber
&&\frac{dF\left( t\right) }{dt}=  \\ \nonumber
&&\allowbreak \Big\{ \left[ \frac{df}{dt}+4if\frac{dg}{dt}-4f^{2}\exp \left( 4ig\right) \frac{dh}{dt}\right] x^{2}+ \\
&&\left[ \frac{dg}{dt}+2if\exp \left( 4ig\right) \frac{dh}{dt}\right] \left(xp+px\right) +\exp \left( 4ig\right) \frac{dh}{dt}p^{2}\Big\}F(t).
\end{eqnarray}
We get back to the original expression for the series
\begin{equation}
F\left( t\right) =\exp \left\{ -t\left[ p^{2}-4x^{2}+2i\left( xp+px\right)\right] \right\} 1
\end{equation}
and taken the derivative with respect to $t$,
\begin{eqnarray}\nonumber
\frac{dF\left( t\right) }{dt}&=&-\left[ p^{2}-4x^{2}+2i\left( xp+px\right)\right] \exp \left\{ -t\left[ p^{2}-4x^{2}+2i\left( xp+px\right)\right] \right\} 1 \\
&=&\left[ -p^{2}+4x^{2}-2i\left( xp+px\right)\right] F.
  \end{eqnarray}
Comparing now both expressions, we get the system of differential equations
\begin{eqnarray}\label{212000}
  \frac{df}{dt}+4if\frac{dg}{dt}-4f^{2}\exp \left( 4ig\right) \frac{dh}{dt} &=& 4, \nonumber \\
  \frac{dg}{dt}+2if\exp \left( 4ig\right) \frac{dh}{dt} &=& -2i, \\
  \exp \left( 4ig\right) \frac{dh}{dt} &=& -1. \nonumber
\end{eqnarray}
The initial conditions that we must set on these equations  and from the ansatz (\ref{ansatz}) are that for $t=0$ the operator in the right side of (\ref{211000}) is the identity, so  we must impose $f(0)=g(0)=h(0)=0$.
The solutions then are simply
\begin{equation}f=\frac{4t}{4t+1},\end{equation}
\begin{equation}g=-\frac{i}{2}\ln \left( 4t+1\right),\end{equation}
\begin{equation}h=-\frac{t}{4t+1}.\end{equation}

We calculate now explicitly
\begin{equation}
F\left( t\right) =\exp \left( fx^{2}\right) \exp \left[ g\left(xp+px\right) \right] \exp \left( hp^{2}\right) 1.
\end{equation}
Remembering  the definition of $p$ (expression \ref{p}) it is very easy to see that
\begin{equation}   \exp \left( hp^{2}\right) 1=1   \end{equation}
and then
\begin{equation}
F\left( t\right) =\exp \left( fx^{2}\right) \exp \left[ g\left(xp+px\right) \right] 1.
\end{equation}
Using now the $\left[ x,p\right] =i$, we write
\begin{equation}
\exp \left[ g\left( xp+px\right) \right] 1=\exp \left[ g\left( 2xp-i\right)\right]
1=\exp \left( -ig\right) \exp \left( 2gxp\right) 1
\end{equation}
and it is also clear that
\begin{eqnarray}
\exp \left( 2gxp\right) 1&&=\left[ 1+gxp+\frac{g^{2}}{2}\left(xp\right) ^{2}+...\right] 1= \\
&&1+gxp1+\frac{g^{2}}{2}\left(xp\right) ^{2}1+...=1\\
 \end{eqnarray}
and also that
\begin{equation}
\exp \left[ g\left( xp+px\right) \right] 1=\exp \left( -ig\right).
\end{equation}
We then have
\begin{equation}
F\left( t\right) =\exp \left( fx^{2}-ig\right).
\end{equation}
Substituting the functions $f$ y $g$,
\begin{equation}
F\left( t\right) =\exp \left[ \left( \frac{4t}{4t+1}\right) x^{2}-\frac{1}{2}\left[
\ln \left( 4t+1\right) \right] \right] =\frac{1}{\sqrt{4t+1}}\exp \left( \frac{4tx^{2}}{4t+1}\right),
\end{equation}
and finally we get the formula we were looking for
\begin{equation} \label{finsumherpar}
\sum\limits_{n=0}^{\infty }\frac{t^{n}}{n!}H_{2n}\left( x\right) =\frac{1}{\sqrt{4t+1}}\exp \left( \frac{4tx^{2}}{4t+1}\right).
\end{equation}
\\

\subsection{Addition formula\index{Hermite polynomials!addition formula}}
We want to apply the form obtained in (\ref{Hermite}), $H_{n}(x)=(-i)^{n}\left( p+2ix\right)^{n}1$, to evaluate the quantity
\begin{equation}\nonumber
H_n(x+y).
\end{equation}
We write it as
\begin{equation}\label{addin}
H_n(x+y)=(-i)^n{[-i\frac{d}{d(x+y)}+2i(x+y)]^n},
\end{equation}
by using the chain rule we have
\begin{equation}
\frac{d}{d(x+y)}=\frac{1}{2}\left(\frac{\partial}{\partial x}+\frac{\partial}{\partial y}\right),
\end{equation}
so that we may re-express (\ref{addin}) in the form
\begin{equation}
H_n(x+y)=\left(\frac{-i}{\sqrt{2}}\right)^n(-i\frac{\partial}{\partial\sqrt{2}x}+2i\sqrt{2}x-i\frac{\partial}{\partial \sqrt{2}y}+2i\sqrt{2}y)^{n}.
\end{equation}
By defining
\begin{equation}
p_X=-i\frac{\partial}{\partial X}, \qquad p_Y=-i\frac{\partial}{\partial Y}
\end{equation}
with $X=\sqrt{2}x$ and $Y=\sqrt{2}y$, we obtain
\begin{eqnarray}
H_n(x+y)=\frac{1}{2^{n/2}}\sum_{k=0}^{n}
\left(
\begin{array}{c}
n\\k
\end{array}
\right)
(-i)^k{(p_X+2iX)^k}(-i)^{n-k} (p_Y+2iY)^{n-k},
\end{eqnarray}
that by using $H_{j}(x)=(-i)^{n}(p+2ix)^{j}1$ adds to
\begin{equation}
H_n(x+y)=\frac{1}{2^{n/2}}\sum_{k=0}^{n}\left(
\begin{array}{c}
n\\k
\end{array}
\right)H_k(\sqrt{2}x)H_{n-k}(\sqrt{2}y),
\end{equation}
which is the addition formula we were looking for.
\section{Associated Laguerre polynomials}
The generating function for the associated Laguerre polynomials\index{associated Laguerre polynomials!generating function} is
\begin{equation}\label{lag1}
\sum_{n=0}^\infty L_n^\alpha(x) t^n = \frac{1}{{(1-t)}^{\alpha+1}} \exp{\left( \frac {-xt}{1-t} \right ) },\qquad |t|<1 .
\end{equation}

The associated Laguerre polynomials may be obtained from the corresponding Rodrigues' formula\index{associated Laguerre polynomials!Rodrigues' formula}
\begin{equation}\label{lag2}
L_{n}^{\alpha }\left( x\right) =\frac{1}{n!}x^{-\alpha}e^{x}\frac{d^{n}}{dx^{n}}\left( e^{-x}x^{n+\alpha }\right).
\end{equation}

The associated Laguerre polynomials satisfy several recurrence relations\index{associated Laguerre polynomials!recurrence relations}. One very useful, when extracting properties of the wave functions of the hydrogen atom, is
\begin{equation}\label{lagrr}
(n+1)L_{n+1}^{\alpha}(x)=(2n+\alpha+1-x)L_{n}^{\alpha}(x)-(n+\alpha)L_{n-1}^{\alpha}(x).
\end{equation}

We will use the operator method outlined above for the Hermite polynomials, to derive the usual explicit expression for the associated Laguerre polynomials. We rewrite expression (\ref{lag2}) as
\begin{equation}
L_{n}^{\alpha }\left( x\right) =\frac{1}{n!}x^{-\alpha}e^{x}\left(ip\right) ^{n}e^{-x}x^{n+\alpha },
\end{equation}
where again the operator $p=-id/dx$, defined in (\ref{p}), is used. \\
We notice that $e^{x}\left(ip\right) ^{n}e^{-x}=[e^{x}\left(ip\right) e^{-x}]^n$ and that, using (\ref{bch0}), $e^xpe^{-x}=(p+i)$, so
\begin{eqnarray*}
L_n^{\alpha}(x)&=&\frac{i^{n}}{n!}x^{-\alpha }\left( p+i\right) ^{n}x^{n+\alpha },
\end{eqnarray*}
or writing explicitly the operator $p$,
\begin{equation}\label{Lag}
L_n^{\alpha}(x)=\frac{1}{n!}x^{-\alpha}\left(\frac{d}{dx}-1\right)^nx^{n+\alpha}.
\end{equation}
Using the binomial expansion,
\begin{equation}
L_n^{\alpha}(x)= \frac{1}{n!}x^{-\alpha}\sum_{m=0}^n\left(
\begin{array}{c}
n\\m
\end{array}
\right)(-1)^{n-m}\frac{d^m}{dx^m}x^{n+\alpha},
\end{equation}
and because
\begin{equation}
\frac{d^m}{dx^m}x^{n+\alpha}=\frac{(n+\alpha)!}{(n+\alpha-m)!}x^{n+\alpha-m},
\end{equation}
we obtain the usual form for associated Laguerre polynomials,
\begin{equation}
L_n^{\alpha}(x)=\sum_{k=0}^n
\left(
\begin{array}{c}
n+\alpha\\n-k
\end{array}
\right)(-1)^k\frac{x^k}{k!}.
\end{equation}
\\

\section{Bessel functions of the first kind of integer order}
Bessel functions of the first kind of integer order, $J_n(x)$, are solutions of the Bessel differential equation\index{Bessel functions of the first kind!differential equation}
\begin{equation} \label{bessel10}
x^2y''+xy'+(x^2-n^2)y=0,
\end{equation}
where $n$ is an integer.
They may be obtained from the generating function\index{Bessel functions of the first kind!generating function}
\begin{equation} \label{bessel20}
\exp\Bigg[{\frac{x}{2}\left(t-\frac{1}{t}\right)}\Bigg]=\sum_{n=-\infty}^{\infty}t^nJ_n(x),
\end{equation}
and also from the following recurrence relations\index{Bessel functions of the first kind!recurrence relations}
\begin{equation}\label{bessel30}
\frac{2n}{x}J_{n}(x)=J_{n-1}(x)+J_{n+1}(x).
\end{equation}

Bessel functions of the first kind of integer order may be written as
\begin{equation} \label{bessel40}
J_n(x)=\sum_{m=0}^{\infty}\frac{(-1)^mx^{2m+n}}{2^{2m+n}m!(m+n)!},
\end{equation}
and also the following integral representation is very useful
\begin{equation}\label{bessel50}
J_{n}\left( x\right) =\frac{1}{2\pi }\int_{-\pi }^{\pi }e^{-i(n\tau-x\sin\tau)}d\tau.
\end{equation}
Some other important relations for the Bessel functions of the first kind are the Jacobi-Anger expansions\index{Bessel functions of the first kind!Jacobi-Anger expansions}:
\begin{equation}\label{ja1}
    e^{ix\cos y}=\sum_{n=-\infty}^{\infty}i^nJ_n(x)e^{iny}
\end{equation}
and
\begin{equation}\label{ja2}
    e^{ix\sin y}=\sum_{n=-\infty}^{\infty}J_n(x)e^{iny}.
\end{equation}

\subsection{Addition formula\index{Bessel functions of the first kind!addition formula}}
Using the operator methods developed in previous sections, we will obtain here the addition formula for the Bessel functions of the first kind of integer order.
\newline
First, we will derive the following expression for any "well behaved" function $f$,
\begin{equation}\label{af1}
 f(x+y)= e^{iyp_x}f(x)e^{-iyp_x}1,
\end{equation}
where $p_x=-id/dx$ is the operator introduced in Section 1, expression (\ref{p}).
Because $e^{-iyp_x}1=1$, developing the $f$ function in a Taylor series (we call $c_n$ to the coefficients in the expansion) and using the linearity of the $e^{iyp_x}$ operator,
\begin{equation}
    e^{iyp_x}f(x)e^{-iyp_x}1=e^{iyp_x}f(x)=e^{iyp_x}\sum_{k=0}^{\infty}c_{k}x^k=\sum_{k=0}^{\infty}c_{k}e^{iyp_x}x^k.
\end{equation}
Now
\begin{equation}
e^{iyp_x}x^k=\sum_{l=0}^{\infty}\frac{(iy)^l}{l!}(-i)^l\frac{d^l}{dx^l}x^k=\sum_{l=0}^{k}(y)^l
\left(
\begin{array}{c}
k\\l
\end{array}
\right)
 x^{k-l}=(x+y)^k,
\end{equation}
then
\begin{equation}
    e^{iyp_x}f(x)e^{-iyp_x}1=\sum_{k=0}^{\infty}c_{k}(x+y)^k=f(x+y),
\end{equation}
as we wanted to prove. \\
Now consider the Bessel function $J_n$ evaluated at $x+y$. From expression (\ref{af1}) we have
\begin{equation}
J_n(x+y)= e^{iyp_x}J_n(x)e^{-iyp_x}1,
\end{equation}
because $e^{-iyp_x}1=1$, and developing the first exponential in Taylor series, we obtain
\begin{equation}\label{forad}
J_n(x+y)= \sum_{m=0}^{\infty}\frac{y^m}{m!}\frac{d^m}{dx^m}J_n(x).
\end{equation}
To calculate the \textit{m}-derivative of $J_{n}$, we use the integral representation (\ref{bessel50}) to write
\begin{equation}
   \frac{d^m}{dx^m}J_n(x) = i^m \frac{1}{2\pi }\int_{-\pi }^{\pi } \sin^m\tau e^{-i(n\tau-x\sin\tau)}d\tau,
\end{equation}
substituting $\sin\tau=(e^{i\tau}-e^{-i\tau})/2i$, and using the binomial expansion,
\begin{eqnarray}\nonumber
   \frac{d^m}{dx^m}J_n(x) &=&
    \frac{1}{2^m} \dfrac{1}{2\pi } \sum_{k=0}^{m} (-1)^k
\left(
\begin{array}{c}
m\\k
\end{array}
\right) \int_{-\pi }^{\pi } e^{i(m-k)\tau} e^{-ik\tau} e^{-i(n\tau-x\sin\tau)}d\tau \\
    &=&\frac{1}{2^m} \frac{1}{2\pi } \sum_{k=0}^{m} (-1)^k \left(
\begin{array}{c}
m\\k
\end{array}
\right) \int_{-\pi }^{\pi }  e^{-i\left[(n-m+2k)\tau-x\sin\tau\right]}d\tau,
\end{eqnarray}
and therefore, using again the integral representation (\ref{bessel50}), we obtain
\begin{equation}
\frac{d^m}{dx^m}J_n(x)=\frac{1}{2^m}\sum_{k=0}^{m}(-1)^k \left(
\begin{array}{c}
m\\k
\end{array}
\right)J_{n-m+2k}(x).
\end{equation}
Substituting this last expression in equation (\ref{forad}), we obtain (we have taken the sum up to infinite as we add only zeros)
\begin{equation}
J_n(x+y)= \sum_{m=0}^{\infty}\frac{y^m}{m!}
\frac{1}{2^m}\sum_{k=0}^{\infty}(-1)^k  \left(
\begin{array}{c}
m\\k
\end{array}
\right)J_{n-m+2k}(x).
\end{equation}
We now change the order of summation and start the second sum at $m=k$ (because from $m<k$ all the terms are zero)
\begin{equation}
J_n(x+y)=\sum_{k=0}^{\infty}\frac{(-1)^k}{k!}
\sum_{m=k}^{\infty}\frac{y^m}{2^m(m-k)!} J_{n-m+2k}(x).
\end{equation}
We do now $j=m-2k$ and obtain
\begin{equation}
J_n(x+y)=\sum_{k=0}^{\infty}\frac{(-1)^k}{k!}
\sum_{j=-k}^{\infty}\frac{y^{j+2k}}{2^{j+2k}(j+k)!} J_{n-j}(x),
\end{equation}
take the second sum from minus infinite, and exchange the order of the sums
\begin{equation}
J_n(x+y)=\sum_{j=-\infty}^{\infty}J_{n-j}(x)\sum_{k=0}^{\infty}\frac{(-1)^k}{k!}
\frac{y^{j+2k}}{2^{j+2k}(m+k)!}=
\sum_{j=-\infty}^{\infty}J_{n-j}(x)J_j(y).
\end{equation}
The final expression
\begin{equation}
J_n(x+y)=\sum_{k=-\infty}^{\infty}J_{n-k}(x)J_k(y)
\end{equation}
is known as the {\it addition formula} for the Bessel functions of the first kind of integer order.
\section{Conclusions}
We have shown how to apply some of the formalism of operator theory to some special functions, namely, Hermite and Laguerre polynomials and Bessel functions.
We have applied them to obtain  some known series of functions, such as the addition formula for Bessel functions, and some, to our knowledge, not known series, such as the
sum of even Hermite polynomials.

\end{document}